\documentclass[aps,pre,reprint,groupedaddress]{revtex4-1}

\usepackage{amsmath,graphicx}
\begin{document}


\title{Parametric decay of oblique Alfv\'en waves in two-dimensional hybrid simulations}

\author{D.~Verscharen}
\email{now at: Space Science Center, Institute for the Study of Earth, Oceans, and Space, University of New Hampshire, Durham, NH 03824, USA; daniel.verscharen@unh.edu}
\affiliation{Max-Planck-Institut f\"ur Sonnensystemforschung, Max-Planck-Str.~2, 37191 Katlenburg-Lindau, Germany}
 \affiliation{Institut f\"ur Theoretische Physik, Technische Universit\"at Braunschweig, Mendelssohnstr. 3, 38106 Braunschweig, Germany}

\author{E.~Marsch}
 \email{marsch@physik.uni-kiel.de}
\affiliation{Max-Planck-Institut f\"ur Sonnensystemforschung, Max-Planck-Str.~2, 37191 Katlenburg-Lindau, Germany}
\affiliation{Institut f\"ur Experimentelle und Angewandte Physik, Christian-Albrechts-Universit\"at Kiel, Leibnizstr. 11, 24098 Kiel, Germany}

\author{U.~Motschmann}
 \email{u.motschmann@tu-braunschweig.de}
\affiliation{Institut f\"ur Theoretische Physik, Technische Universit\"at Braunschweig, Mendelssohnstr.~3, 38106 Braunschweig, Germany}
\affiliation{Institut f\"ur Planetenforschung, DLR, Rutherfordstr.~2, 12489 Berlin-Adlershof, Germany}

\author{J.~M\"uller}
 \email{joa.mueller@tu-bs.de}
\affiliation{Institut f\"ur Theoretische Physik, Technische Universit\"at Braunschweig, Mendelssohnstr.~3, 38106 Braunschweig, Germany}

\date{July 19, 2012}

\begin{abstract}
Certain types of plasma waves are known to become parametrically unstable
under specific plasma conditions, in which the pump wave will decay into
several daughter waves with different wavenumbers and frequencies. In the
past, the related plasma instabilities have been treated analytically for
various parameter regimes and by use of various numerical methods, yet the
oblique propagation with respect to the background magnetic field has rarely
been dealt with in two dimensions, mainly because of the high computational
demand. Here we present a hybrid-simulation study of the parametric decay of
a moderately oblique Alfv\'en wave having elliptical polarization. It is
found that such a compressive wave can decay into waves with higher and lower
wavenumbers than the pump.
\end{abstract}


\maketitle

\section{Introduction}

Monochromatic plasma waves with certain properties are known to be
parametrically unstable and to decay into daughter waves in a multiple-waves
interaction process
\cite{galeev63,derby78,goldstein78,lashmore-davies79,wong86,inhester90,ruderman04}.
The ubiquitous small-amplitude thermal fluctuations existing in any plasma
are considered as seeds for growing daughter waves in the presence of a
large-amplitude pump wave, if that has the necessary characteristics and
fulfills the required instability criteria.
Multiple-wave interactions provoke these instabilities. This fact directly illustrates that they are nonlinear processes by nature. The monochromatic initial condition makes the parametric decay an illustrative example compared to other nonlinear mechanisms. Therefore, it is of general interest for a better understanding of more intricate nonlinear effects in plasmas and other statistical systems ranging from classical fusion \cite{gnavi96,baldis97} and space plasmas \cite{araneda07,tanaka07} to more exotic media such as relativistic electron-positron plasmas \cite{gomberoff97,matsukiyo03}.

Following the early analytical treatments, modern numerical simulations are
capable of modeling the parametric decay \cite{vinas91a,araneda98}
comprehensively. Kinetic simulations even allow one to investigate in detail
the interactions between the participating particles and waves, and so can
reveal in particular the resultant particle heating, e.g., under conditions
typical for the solar corona holes \cite{araneda08}. This possibility has
brought the parametric decay process into the focus of coronal-heating
research. In the past decade, large-amplitude magnetohydrodynamic waves have
also been observed directly in the solar chromosphere and corona
\cite{nakariakov05}. They seem to be mainly Alfv\'enic, yet with a smaller
slow-mode-wave compressive component, and appear intense enough to deliver
via dissipation sufficient thermal energy to the coronal ions
\cite{depontieu07,mcintosh11}. This makes them a promising energy source also
for the acceleration of the fast solar wind, even though the details of the
dissipation and the spectral transfer are not well understood
\cite{marsch06,cranmer09,ofman10}.

Most models made use of simplifications such as one-dimensional geometry. But
recently, the more powerful available computers have paved the way for fully
two-dimensional analyzes, including the possibility of oblique propagation of
the mother and daughter waves \cite{vinas91b,matteini10}. In this context
there remain, however, many open physical questions. The compressive
component of the fluctuations, for example, is long known to be important for
the parametric decay. Yet in the oblique case, the effects of compressibility
are not well analyzed or understood. Also, the direction of propagation of
the daughter-wave products and their ability for resonant wave--particle
heating are still unclear. This work will address some of these aspects and
issues with the aid of numerical hybrid simulations. The treated plasma gains further degrees of freedom by choosing higher dimensionality. Since natural systems are never limited to a one-dimensional geometry only, it is crucial to understand the consequences of this advancement towards a more realistic modeling. Further complications have been treated in simplified cases before, such as the influence of additional ionic species  \cite{mishra93,araneda09}, dusty components in the plasma \cite{amin96,hertzberg04,shukla06}, and high collision rates \cite{vladimirov93}. The present work is supposed to provide a basis for further studies of parametric instabilities under the refined conditions adapted to the particular situation.

\section{Numerical method}

The so called A.I.K.E.F. (adaptive ion-kinetic electron-fluid) code is a
numerical hybrid code, which treats ions as particles following the
characteristics of the Vlasov kinetic equation and electrons as a massless
charge-neutralizing fluid. Hybrid codes have been successfully applied  in the past  to model plenty of plasma phenomena in the limit of low frequencies compared to any frequency related to the electron dynamics \cite{winske96}. The dynamics of ions are appropriately simulated as long as quasi-neutrality is satisfied. These constraints are often fulfilled in space plasmas. Therefore, hybrid simulations have proven to be adequate models for effects such as micro-instabilities \cite{daughton99,gary03,cowee10}, the turbulent cascade \cite{markovskii09,verscharen12,vasquez12}, or micro-physics at planetary magnetospheres \cite{kallio11,mueller12} to name but a few examples. 

The A.I.K.E.F. code has been described by M\"uller et al.
\cite{mueller11} and already been used for a treatment of waves and
turbulence in space plasmas \cite{verscharen12}. The equations of motion
solved for a proton are
\begin{align}
\frac{\mathrm d\vec v_{\mathrm p}}{\mathrm d t}&=\frac{q_{\mathrm p}}{m_{\mathrm p}}\left(\vec E+\frac{1}{c}\vec v_{\mathrm p}\times \vec B\right), \label{eqmotion_fla}\\
\frac{\mathrm d\vec x_{\mathrm p}}{\mathrm dt}&=\vec v_{\mathrm p},
\end{align}
with the velocity $\vec v_{\mathrm p}$ and the spatial location vector $\vec
x_{\mathrm p}$. We consider the Lorentz force which acts on any particle with
charge $q_{\mathrm p}$ and mass $m_{\mathrm p}$, and is due to the electric
field $\vec E$ and the magnetic field $\vec B$. The speed of light is denoted
by $c$.

Concerning the electromagnetic fields involved, we use the momentum equation
of the massless electron fluid to deliver the electric field as
\begin{equation}
\vec E=-\frac{1}{c}\vec u_{\mathrm e}\times \vec B-\frac{1}{n_{\mathrm e}e}\nabla p_{\mathrm e},
\end{equation}
where the electron bulk velocity is denoted by $\vec u_{\mathrm e}$, the
number density by $n_{\mathrm e}$, and its elementary charge by $e$. The
magnetic field is obtained from the induction equation, which means from
Faraday's law,
\begin{equation}
\partial \vec B/\partial t=-c \nabla \times \vec E.
\end{equation}

The electrons are assumed to be isothermal, and thus their pressure
$p_{\mathrm e}$ depends on the electron number density according to
$p_{\mathrm e}\propto n_{\mathrm e}$. The proton density and the proton bulk
velocity $\vec u_{\mathrm p}$ are obtained as the first two moments of the
proton distribution function, and quasi-neutrality connects the density of
protons and electrons.

The boundaries of the simulation box are set to be periodic, and the
particles are initialized with a Maxwellian velocity distribution that is
shifted to the given values for the initial bulk velocities and has a width
determined by the proton beta, which is set to $\beta_{\mathrm
p}=\beta_{\mathrm e}=0.1$. All spatial length scales are normalized and given
in units of the proton inertial length, $\ell_{\mathrm p}=c/\omega_{\mathrm
p}$, with the proton plasma frequency $\omega_{\mathrm p}=\sqrt{4\pi
n_{\mathrm p}q_{\mathrm p}^2/m_{\mathrm p}}$. All time scales are in units of
the inverse proton gyro-frequency $\Omega_{\mathrm p}=q_{\mathrm
p}B/(m_{\mathrm p}c)$.

The two-dimensional integration box has a size of $1024\times 1024$ cells.
Each cell is filled with 500 super-particles, which represent the real number
of the protons. For the vector quantities, all three components are
evaluated, and a divergence-cleaning algorithm is applied to guarantee
numerical stability. A constant background magnetic field of the form $\vec
B_0=B_0  {\vec e}_z$ is set up, and $B_0$ provides the normalization unit
for all magnetic fields.

The initial condition is a monochromatic 
Alfv\'en/ion-cyclotron (A/IC) wave with a normalized amplitude of $b=0.2$.
Its polarization is determined by the Hall-MHD relations, since this is the
low-temperature limit of the hybrid equations \cite{vocks99}. The wave
propagates with an angle of $\vartheta=10^{\circ}$ with respect to the
background magnetic field. Such a wave is elliptically polarized and has an
initially compressive component. The spatial domain has a size of roughly
$250\ell_{\mathrm p}\times 250\ell_{\mathrm p}$, restricted by the periodic
connection on the boundaries.

\section{Results}

\begin{figure}
\includegraphics[width=\columnwidth]{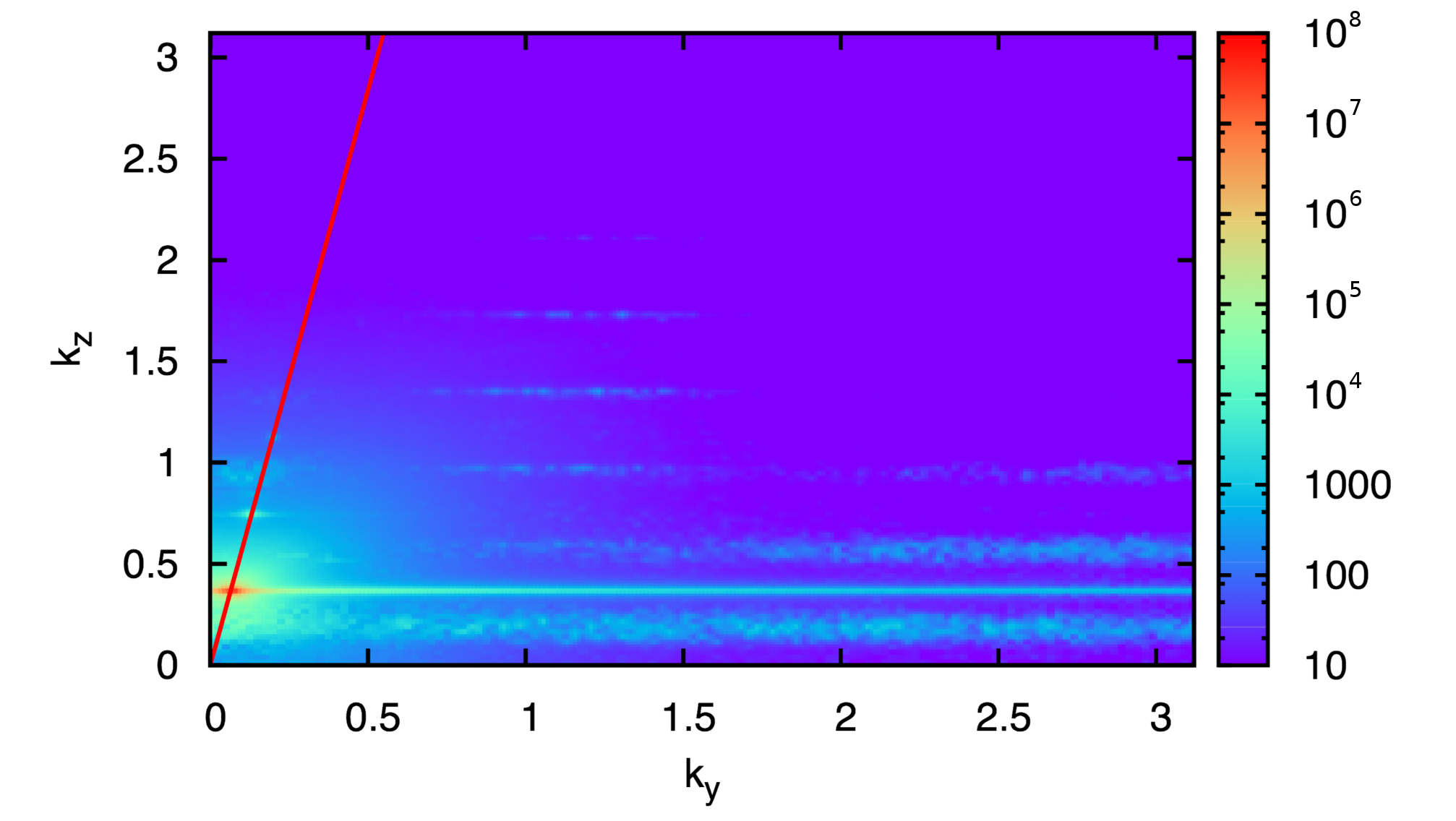}%
\caption{(Color online) Two-dimensional power spectral density of the magnetic field
fluctuations at $t=500$. The straight line starting at the origin (red) indicates the initial propagation
direction with $\vartheta=10^{\circ}$. The color coding represents the power
spectral density in arbitrary units.}
\label{fig_paramdec_ob_spec}
\end{figure}

After 50\,000 time steps, corresponding to the time $t=500$, the initial wave
has decayed already. The resulting two-dimensional spectrum is shown in
Fig.~\ref{fig_paramdec_ob_spec}. The initial oblique pump wave is still
visible as a bright dot at $k_y\approx 0.06$ and $k_z\approx 0.38$.
Apparently, the wave energy has mainly been transported along the initial
propagation direction of the wave. It seems to have a higher power up to
$k_z=1$ in normalized units, which means that here a spectral break should be
expected. There is an increased wave activity at higher $k_y$, and
significant additional power is distributed there. At low values of
$k_z$, two broader perpendicular patterns are visible in the side-bands of the pump.

\begin{figure}
\includegraphics[width=\columnwidth]{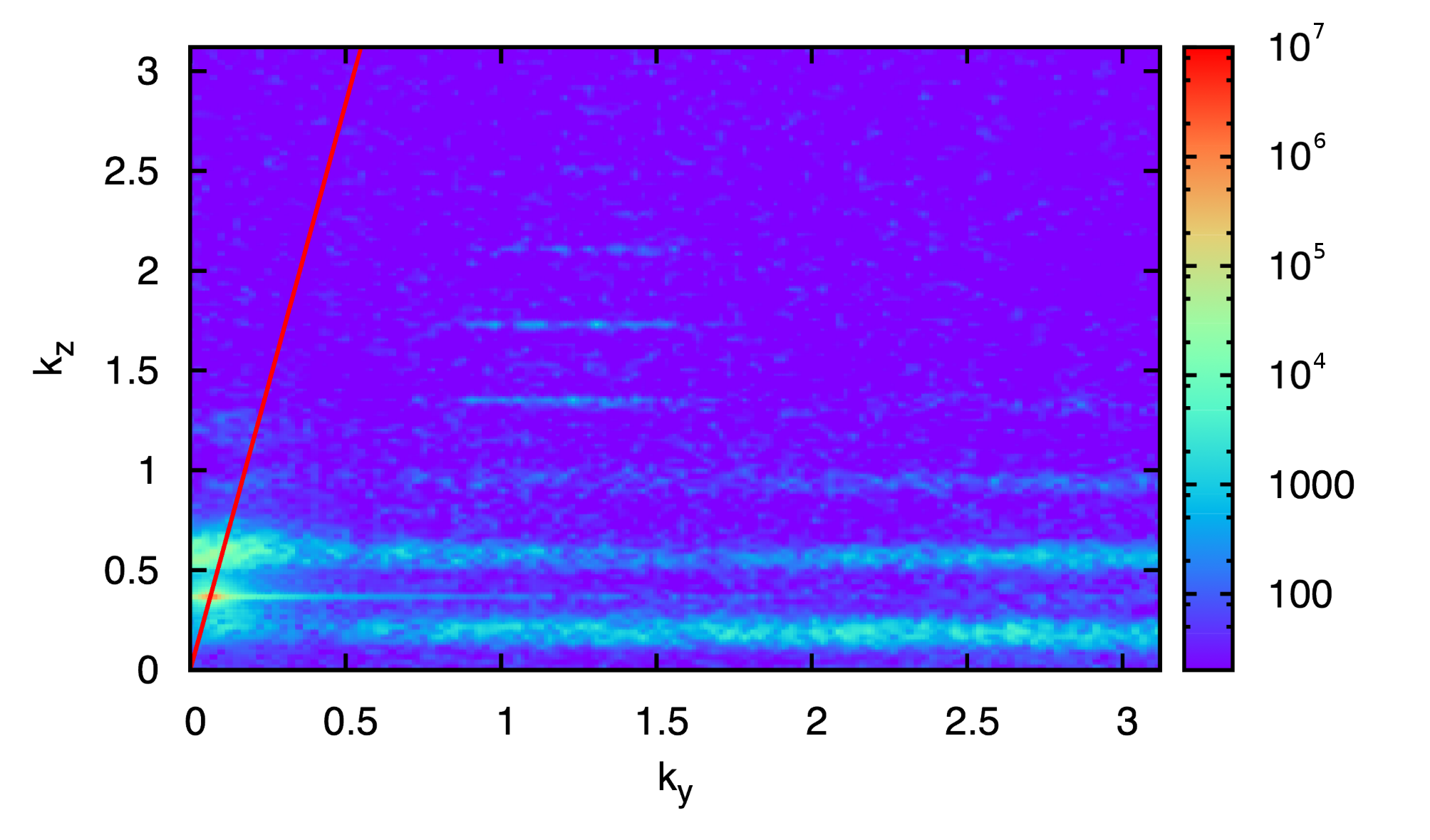}%
\caption{(Color online) Two-dimensional power spectral density of the density fluctuations
at $t=500$. The straight line starting at the origin (red) indicates the initial propagation direction
with $\vartheta=10^{\circ}$. The color coding represents the power spectral
density in arbitrary units.}
\label{fig_paramdec_ob_density}
\end{figure}

Also, the density fluctuation spectrum in two dimensions can be calculated
from the simulation data. It is shown in Fig.~\ref{fig_paramdec_ob_density},
which indicates that the main features visible in the magnetic fluctuation
spectrum also appear in that of the density, and thus the waves posses
compressive components. This is especially true for all components having a
non-zero $k_y$. Owing to the lower overall level of these fluctuations, some filamentary intermediate structures are distinctly visible. These should be understood as representing the broad-band compressive component of the daughter waves. They occur at higher harmonics of the first side-bands, with spacing given by the parallel wavenumber of the pump wave.

\begin{figure}
\includegraphics[width=\columnwidth]{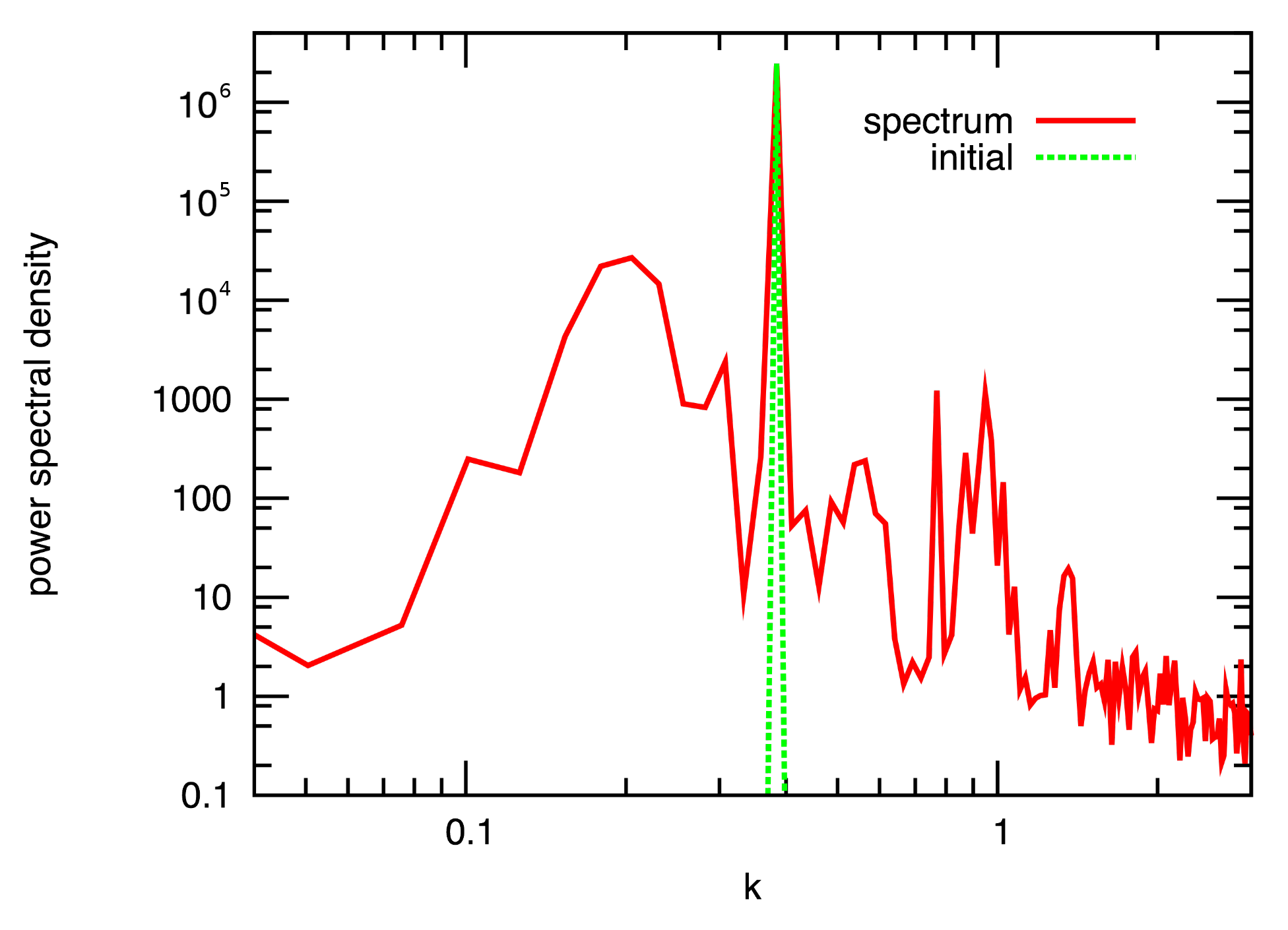}%
\caption{(Color online) One-dimensional power spectral density for the magnetic field
fluctuations at $t=500$ along the initial direction of propagation.
Additionally, the initial spectrum is shown (green dashed line).}
\label{fig_paramdec_ob_1d_B}
\end{figure}

\begin{figure}
\includegraphics[width=\columnwidth]{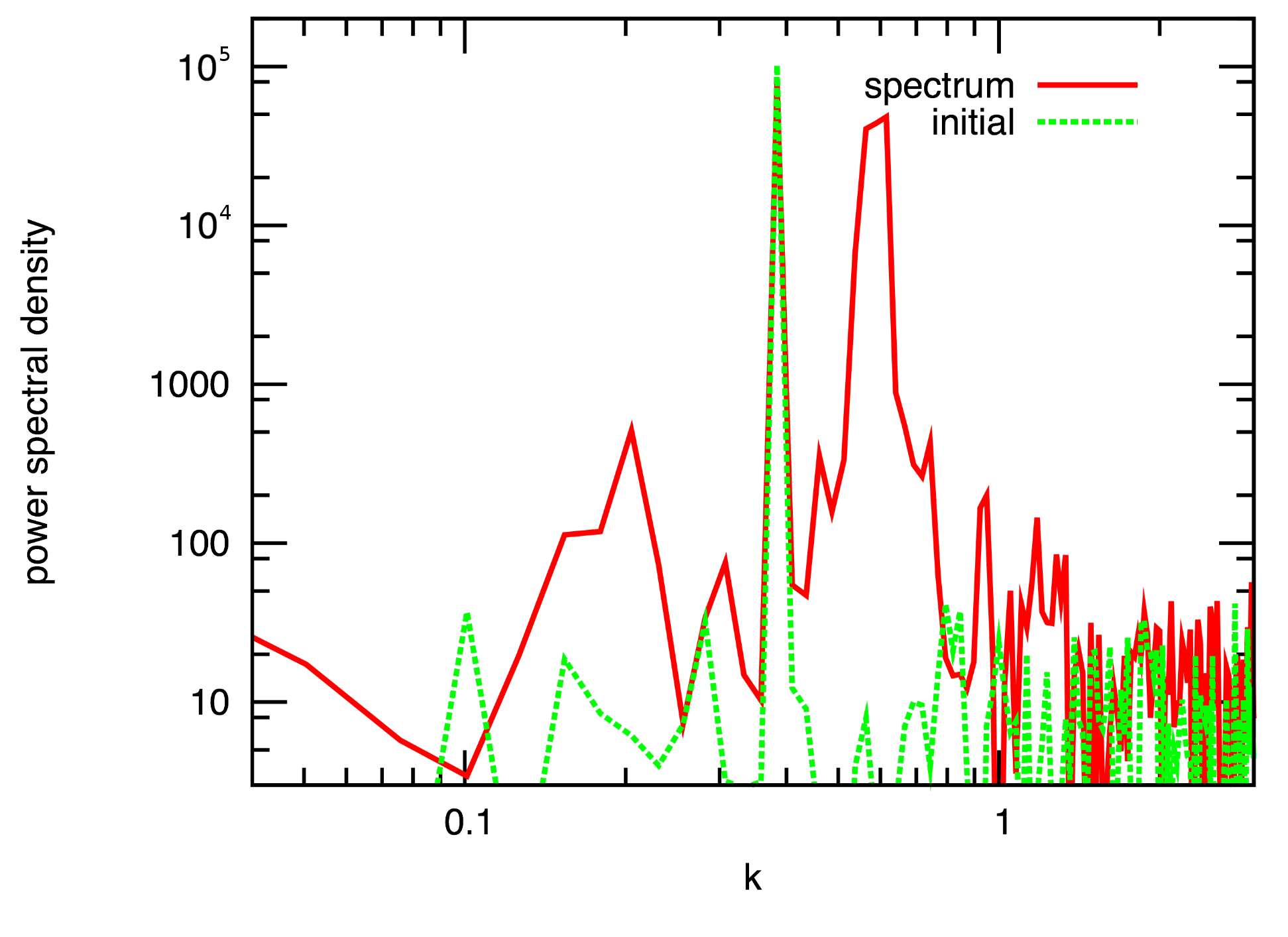}%
\caption{(Color online) One-dimensional power spectral density for the density fluctuations
at $t=500$ along the initial direction of propagation. Additionally, the
initial spectrum is shown (green dashed line).}
\label{fig_paramdec_ob_1d_rho}
\end{figure}

To study the power distribution in more detail, the 2D Fourier transform can
be cut and displayed along the direction of the initial propagation. It
corresponds to a cut along the red lines in Figs.~\ref{fig_paramdec_ob_spec}
and \ref{fig_paramdec_ob_density}. The corresponding one-dimensional power
spectra are shown in Figs.~\ref{fig_paramdec_ob_1d_B} and
\ref{fig_paramdec_ob_1d_rho}.

The initial wave is still visible in both the magnetic field fluctuations and
the density fluctuations. Remember that the initial wave was compressive
already, due to its oblique propagation implying compressibility. The initial
wave looses some energy as compared to the beginning energy. It is first
spread to different wavenumbers and then dissipated at the small kinetic
scales. Thus the pump wave decays to daughter waves with higher and lower
wavenumbers in comparison to the initial ones.

From the simulation, the dispersion of the daughter waves can in general be
determined along any direction in the $(k_y,k_z)$-plane. Since an enhancement
of energy is seen along the initial direction of propagation (i.e., along the
red line in Fig.~\ref{fig_paramdec_ob_spec}), it is appropriate to calculate
the dispersion along this line. Therefore, a two-dimensional spatial Fourier
transform is applied, and a one-dimensional cut is taken along the direction
$\vartheta=10^{\circ}$ for 60 different time steps. They are separated by a
time difference of $1/\Omega_{\mathrm p}$ and thus allow for a temporal
Fourier transformation to be performed.

\begin{figure}
\includegraphics[width=\columnwidth]{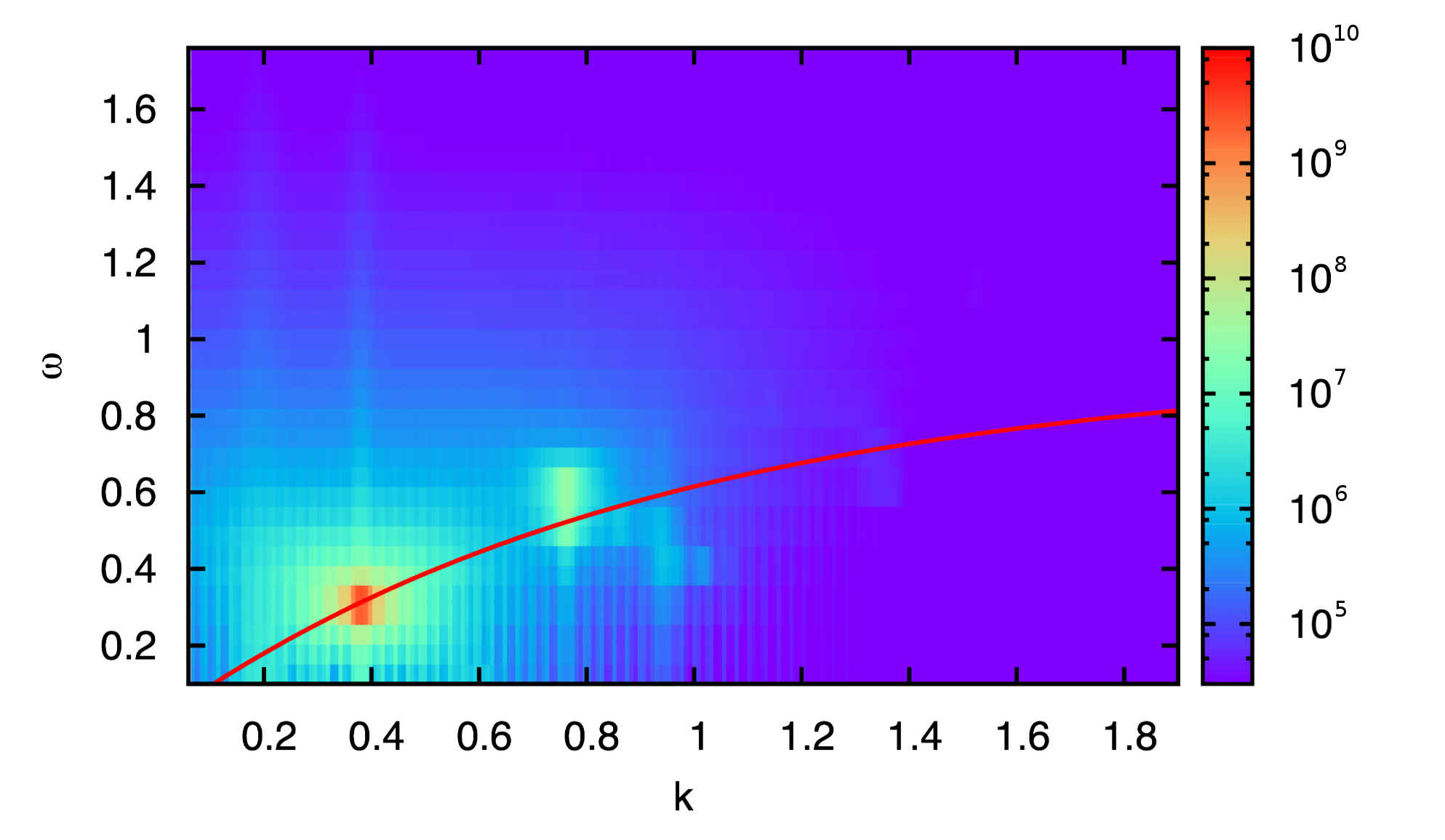}%
\caption{(Color online) One-dimensional dispersion analysis of the magnetic field fluctuations
along the direction $\vartheta=10^{\circ}$ at $t=500$. The color coding represents the
power spectral density. The enhancement in power corresponds to A/IC waves
propagating obliquely to the background magnetic field. The curved solid line starting at the origin (red) shows
the cold-plasma dispersion relation for oblique A/IC waves with $\vartheta=10^{\circ}$.}
\label{fig_paramdec_ob_dispersion}
\end{figure}

The result of the spectral analysis is shown in the dispersion plot of
Fig.~\ref{fig_paramdec_ob_dispersion}. The enhanced power in the
$(k_y,k_z)$-analysis is also very well located in the $(\omega,k)$-plane at
$k\approx 0.75$ and $\omega \approx 0.6$. Comparing this point with the
theoretical cold dispersion analysis of oblique A/IC waves \cite{stix92},
shown as a red curve, reveals that this point is located very close to the
branch of this wave mode. The width of the signal in this diagram is due to
the pump wave, which has a significant amplitude compared to the background
magnetic field. Therefore, waves other than the pump wave encounter an inhomogeneous guiding field due to the magnetic field of the pump wave, being superposed on the background field. The resulting total field can be understood as an effective magnetic field, around which the daughter waves propagate. This effect leads to a small broadening and shift in the dispersion
analysis.

Calculations with further, moderately oblique propagation angles provide
comparable results, especially the outcome that the preferred direction of
the daughter wave propagation is along the initial pump wave direction.

\section{Discussion and Conclusions}

The daughter waves, which are generated already after quite a short time of
the nonlinear evolution, are found to be mainly aligned along the initial
direction of pump wave propagation. This can be understood as a consequence
of the conservation of wave momentum. In general, the wavevectors of the pump
wave and the two daughter waves have to form a vector triangle to fulfill
this conservation in a three-wave process. Other arbitrary combinations would
be possible for an interaction between four or even more waves
\cite{davidson72}. However, it seems that the background magnetic field is
not the most important guiding vector, but rather the initial wave
propagation vector, which determines the geometry of the daughter wave vector
system. The oblique-wave hybrid simulations performed by Matteini et al.
\cite{matteini10} can also be interpreted in this sense, even though the
authors favor the interpretation of a field-parallel spectral transfer.
However, a major difference to their setup is our initialization with an
elliptically polarized Hall-MHD pump wave, apart from the higher numerical
resolution employed. The daughter waves with lower wavenumbers seem to orient
themselves more perpendicular to the background field. The modulational
instability can be understood as the generation mechanism here and seems to
favour this direction of propagation.

An oblique wave is intrinsically compressive due to its polarization. During
its evolution and decay, it is thus prone to generate a broad spectrum of
compressive fluctuations. It is important to remember, though, that the
amplitude of an oblique wave is in any case not arbitrary. Its intrinsic
compressive effects pose an upper limit, because negative density values have
to be avoided and are forbidden \cite{yoon11}.

The dispersion relation of the decay products obtained from the oblique
two-dimensional simulation shows that the related daughter waves are still
A/IC waves, yet with higher wavenumber and frequency. Other wave modes such
as the fast/whistler branch for example do not appear. This result is in
agreement with previous treatments of the parametric decay, which have shown
that the A/IC wave is a typical and major daughter product of the decay
\cite{araneda07}. The initial oblique wave with wavenumber $k_0$ is prone to
the decay instability with $k>k_0$ and the modulational instability with
$k<k_0$. Dispersive effects let the decay instability generate A/IC waves
\cite{hollweg94}. Obviously, this effect can occur very efficiently for
$k\gtrsim 0.6$. Generally, daughter waves are only excited in certain limited
ranges but not on a broad range of wavenumbers. The A/IC waves, however, can
easily fulfill the condition of cyclotron resonance for sufficiently high
frequencies/wavenumbers. Thus, there is an upper limit for the occurrence of
A/IC waves, essentially due to the onset of cyclotron damping. This leads to
quite a sudden cut in the spectrum at $k\approx 0.9$. The dispersion diagram
together with this typical onset of damping underlines the A/IC nature of the
daughter waves.

The temperature of the particles does not increase significantly over the
integration time in our simulations. This may be due to a comparably low
intensity of the daughter waves, and to the limited simulation time.
One-dimensional simulations have shown an increase and saturation of the
particle temperatures, owing to heating by resonant wave--particle
interactions \cite{araneda09}. The temperature in the above simulations does,
however, increase if no divergence-cleaning algorithm is applied to the
magnetic fields. So, maybe a possible heating is suppressed by the present
schemes, or the heating observed in previous simulations is mainly a
numerical artifact. This question cannot be answered conclusively here.

\begin{acknowledgments}
D.~V.~received financial support from the International Max Planck Research
School (IMPRS) on Physical Processes in the Solar System and Beyond. The
calculations have been performed on the MEGWARE Woodcrest Cluster at the
Gesellschaft f\"ur wissenschaftliche Datenverarbeitung mbH G\"ottingen
(GWDG).
\end{acknowledgments}

%

\end{document}